# Indirect bandgap of hBN-encapsulated monolayer MoS$_2$

Yosuke Uchiyama[1], Kenji Watanabe[2], Takashi Taniguchi[2], Kana Kojima[3], Takahiko Endo[3], Yasumitsu Miyata[3], Hisanori Shinohara[1] and Ryo Kitaura[1]*

[1]*Department of Chemistry, Nagoya University, Nagoya 464-8602, Japan*
[2]*National Institute for Materials Science, 1-1 Namiki, Tsukuba 305-0044, Japan*
[3]*Department of Physics, Tokyo Metropolitan University, Hachioji, Tokyo 192-0397, Japan*

*Corresponding authors: Ryo Kitaura, r.kitaura@nagoya-u.jp

**Abstract**

We present measurements of temperature dependence of photoluminescence intensity from monolayer MoS$_2$ encapsulated by hexagonal boron nitride (hBN) flakes. The obtained temperature dependence shows an opposite trend to that of previously observed in a monolayer MoS$_2$ on a SiO$_2$ substrate. Ab-initio bandstructure calculations have revealed that monolayer MoS$_2$ encapsulated by hBN flakes have no longer a direct-gap semiconductor but an indirect-gap semiconductor. This is caused by orbital hybridization between MoS$_2$ and hBN, which leads to upward shift of Γ-valley of MoS$_2$. This work shows an important implication that the hBN-encapsulated structures used to address intrinsic properties of two-dimensional crystals can alter basic properties encapsulated materials.

Recently appearing two-dimensional (2D) materials, including graphene, phosphorene, transition metal dichalcogenides (TMDs), etc., have opened up a new field in the science of low-dimensional materials[1-6]. TMDs, in particular, provide a wide variety of 2D layered materials with various compositions and electronic structures, giving us a widespread and excellent field for exploration of physics in the realm of the 2D world. In contrast to graphene, semiconducting 2D-TMDs can have a sizable bandgap up to ~2 eV, which offers an opportunity to explore optical responses at the 2D limit and develop TMD-based nanoelectronic devices[5,7,8]. Furthermore, 2D-TMDs afford vertical or lateral heterostructures, whose electronic structure and physical properties can be tuned through selecting the combination and stacking angles of each layer. Coupled with the possibility arising from the valley degree of freedom[9,10], TMDs have been yielding new perspectives and attracting a wide range of research interests.

For exploration of the fascinating opportunities, one of the important things is to address intrinsic properties of TMDs. For this purpose, TMDs encapsulated by hexagonal boron nitride (hBN), hBN/TMD/hBN, have been widely used[11-14]. 2D-TMDs are very sensitive to the external environment, such as substrates and adsorbents, because almost all atoms in a 2D-TMD locate at the surface. $SiO_2$/Si usually used as a substrate has a rough surface with dangling bonds, low-energy optical phonons, and charged impurities, which can significantly degrade the quality of samples[15]. In contrast, hBN, a graphene analogue insulator (bandgap ~ 6 eV), is free from these degradation factors, providing an ideal environment to address intrinsic properties of 2D-TMDs.

Up to now, quite a few studies have been done with hBN/TMD/hBN to investigate intrinsic properties of 2D-TMDs. For example, high-mobility devices with hBN-encapsulated structures have been reported, showing carrier mobility of 19-94 $cm^2$/Vs at room temperature in a monolayer $MoS_2$ (ML-$MoS_2$) encapsulated by hBN flakes; typical mobilities of ML-$MoS_2$ on silicon substrates range from 1 to 10 $cm^2$/Vs[16,17]. The enhancement in carrier mobility arises from suppression of the extrinsic carrier scatterings in hBN-encapsulated samples. The high quality of hBN-encapsulated samples has also been observed in optical measurements. Photoluminescence (PL) spectra of a ML-TMD on a silicon substrate each show a relatively broad peak arising from radiative recombination of excitons; for example, monolayer $WS_2$ on a silicon substrate shows a PL peak whose full width at half maximum (FWHM) is typically 50-55 or 75 meV. In contrast, the PL spectrum of a hBN/$WS_2$/hBN shows a corresponding PL peak with a much smaller FWHM of ~26 meV, and this small FWHM mainly results from suppression of inhomogeneous broadening arising from substrates[18-20]. These results strongly indicate that hBN-encapsulated samples are essential to address intrinsic properties of 2D-TMDs.

In this work, we have focused on the electronic structure of one of the most popular TMDs, ML-MoS$_2$, encapsulated by hBN, hBN/MoS$_2$/hBN. As discussed above, hBN-encapsulated structures are probably the best structure for investigations of intrinsic properties of TMDs, but a question here is "does the hBN-encapsulation really preserve the original electronic structure of a 2D-TMD or not?" In multi-layer systems assembled through van der Waals (vdW) interaction, it has been reported that the band structure of vdW stacks can be modulated by inter-layer interaction. Bilayer graphenes are one of the most significant examples. A recent work has revealed that interlayer interaction causes a flat band in a bilayer graphene, which leads to the Mott insulating state and even superconductivity at low temperature. Interlayer interaction in hBN/TMD/hBN, at first sight, is not the case, because hBN have a large bandgap of ~6 eV and the valence band maximum (VBM) and conduction band minimum (CBM) of TMD locate away from those of hBN, but is this really the case?

In this paper, we show that modification of the band structure of ML-MoS$_2$ occurs through interlayer interaction between ML-MoS$_2$ and hBN flakes. Through detailed PL measurements and first-principles band-structure calculations, we have found that the valence band (VB) at the Γ-point of ML-MoS$_2$ shows upshift due to orbital hybridization and structural distortion, which leads to transition from a direct to an indirect semiconductor. A direct gap is one of the most significant features of ML-MoS$_2$; however, in the case of hBN/MoS$_2$/hBN, ML-MoS$_2$ does not have a direct gap anymore, which results in a low-energy momentum-forbidden dark state in optical excitations. This direct-to-indirect gap transition is not sensitive to the relative angle between MoS$_2$ and hBN, altering significantly optical responses of MoS$_2$. This has important implications for investigation of valley-related optical responses at low temperature.

## Results

ML-MoS$_2$ flakes were grown on exfoliated hBN flakes on a quartz substrate with the chemical vapor deposition (CVD) method using MoO$_3$ and elemental sulfur as precursors; a high growth temperature of 1100 degrees Celsius was applied to improve the crystallinity of grown ML-MoS$_2$. A flake of hBN with a thickness of ~15 nm was then transferred onto a ML-MoS$_2$ crystal grown on a hBN flake. After the transfer, the nano-"squeegee" technique[21] was used to remove contaminations encapsulated between the ML-MoS$_2$ and the hBN flakes. Figure 1 shows an optical microscope image of the ML-MoS$_2$ encapsulated by hBN flakes. No bubble is visible in the optical image, indicating that the ML-MoS$_2$ contacts hBN flakes well to form a high-quality hBN-encapsulated structure. An AFM height image of the hBN/MoS$_2$/hBN (Fig. 1(b)) shows an atomically flat surface, giving a height of MoS$_2$ as ~0.8 nm that is consistent with a monolayer structure. The RMS roughness evaluated at the ML-MoS$_2$ region is ~0.1 nm, which clearly demonstrates the atomically flat surface of the prepared sample. This atomically flat structure is essential to suppress inhomogeneous broadening in observations of optical responses.

Figure 2(a) shows a PL spectrum of the prepared hBN/MoS$_2$/hBN measured at 300 K. A single peak arising from radiative recombination of excitons is seen in the PL spectrum, where the contribution from trions is not dominant (a detailed spectral decomposition is shown in Supplementary Figure 1). The FWHM of the exciton PL peak is 39 meV, which is much smaller than those of samples on silicon substrates (typically ~70 or 56 meV)[22,23]. Figure 2(b) shows a PL image of the sample, where the red-dashed rectangle corresponds to the place cleaned by the nano-"squeegee" technique. The PL image, in particular at the red rectangle, clearly shows uniform PL, which represents a high-quality clean interface between ML-MoS$_2$ and hBN. The brighter PL at the cleaned place indicates that non-radiative decay, which is caused by contaminants, is suppressed[19,24,25].

Figure 3(a) shows PL spectra measured at temperatures ranging from 220 to 320 K. PL peaks arising from radiative recombination of excitons show blue shift as temperature decreases, which originates from bandgap widening caused by the electron-phonon interaction. The temperature dependence is well described by Varshni's equation, which is shown in Supplementary Figure 2; the obtained parameters are consistent with previously reported values. As you can clearly see, intensities of the PL peaks become weak as temperature decreases. To evaluate the intensity decrease precisely, we measured PL intensity from the red rectangular regions in PL images; PL intensities are averaged over the area to minimize the position-dependent fluctuation. As clearly seen in Figure 3(b), PL intensity monotonically decreases as temperature decreases, and PL intensity at 260 K is two-thirds of that at 320 K.

In a previous report, PL intensity of ML-MoS$_2$ on a silicon substrate increased as temperature decreased[26]. Ab-initio band structure calculation tells us that ML-MoS$_2$ is a direct-gap semiconductor, and bright excitons might be the lowest-energy excited state. This is consistent with the previously reported temperature dependence of PL intensity because the population of bright excitons is expected to increase as temperature decreases if the bright state is the lowest-energy state. A recent theoretical investigation, however, has suggested that lower-energy dark excitons can exist in a ML-MoS$_2$, which dark excitons correspond to holes located at the Γ-valley[27]; excitons with Γ-valley holes are dark state because direct recombination is forbidden due to the momentum conservation. The momentum-forbidden dark excitons can have lower energy than K-K direct excitons because of a difference in exciton binding energy. Even with the lower-energy dark excitons, PL intensity in MoS$_2$/SiO$_2$ can still increase as temperature decreases because population of another type of dark excitons, K-K direct excitons with a center-of-mass momentum exceeding the light cone, decreases as temperature decreases.

Figure 4 shows the temperature dependence of time-dependent PL intensity measured with the time-correlated single-photon counting technique. As clearly seen, PL decay becomes faster as temperature becomes lower. Coupled with the observed decrease in PL at low temperature, it is strongly suggested that there is a dark state, whose energy is lower than the bright state, in hBN/MoS$_2$/hBN. This dark state should have lower energy than the momentum-forbidden dark states observed in MoS$_2$/SiO$_2$ because temperature dependence in PL intensity of hBN/MoS$_2$/hBN is opposite to that of MoS$_2$/SiO$_2$. To extract the energy difference between bright and dark excitons in the hBN/MoS$_2$/hBN, we have fitted the temperature dependence in Fig. 3(b) with the following equation.

$$I(T) \propto \frac{\alpha m_{KK}}{m_{KK} + m_{KK'} + m_{K\Gamma} e^{\beta \Delta E_{K\Gamma}}}$$

In this equation, temperature-dependent change in the population of spin-forbidden and momentum-forbidden dark excitons is considered; there are two momentum-forbidden dark excitons, inter-valley indirect excitons and direct excitons with momentum exceeding the light cone. As shown by the dashed line in Fig. 3(b), the equation reproduces the observed temperature dependence well, giving $\Delta E_{K\Gamma}$ of 83 meV. This is much larger than the previously reported value for ML-MoS$_2$ on a silicon substrate, which strongly indicates that the electronic structure is modified in hBN/MoS$_2$/hBN.

To address the origin of the dark state in hBN/MoS$_2$/hBN, we have performed ab-initio density functional band structure calculations. Figure 5(a) shows the band structure of isolated ML-MoS$_2$ and ML-MoS$_2$ sandwiched by monolayer hBN. Prior to calculation of hBN/MoS$_2$/hBN, we performed full structural optimization of isolated ML-MoS$_2$, and the optimized structure (optimized primitive vectors and geometry) was used in the calculation of hBN/MoS$_2$/hBN; we used a 4 x 4 and 5 x 5 supercells for MoS$_2$ and hBN to minimize a difference in unit cells. Interlayer distance between MoS$_2$ and hBN in hBN/MoS$_2$/hBN was evaluated with the vdW-DFT method (supplementary Fig. S3). As you can see, CBM and VBM locate at the K-valley in ML-MoS$_2$, which is consistent with the band structure previously reported. On the other hand, the valence band at the Γ-valley shows upward shift and the VBM locates at the Γ-valley in the hBN/MoS$_2$/hBN; i.e. hBN/MoS$_2$/hBN is an indirect semiconductor. Because we used the optimized primitive vectors of MoS$_2$ for the calculation of hBN/MoS$_2$/hBN, the observed change in bandstructure should be caused not by unrealistic strain arising from difference in unit cells but by interlayer interaction between hBN and MoS$_2$. The direct-to-indirect change in bandstructure is consistent to a previous DFT calculation.

This upward shift is very sensitive to interlayer distance between MoS$_2$ and hBN, showing exponential decay against the interlayer distance (Supplementary Fig. S4). This strong interlayer dependence indicates that the upward shift in the VBM at the Γ-valley originates not from long-range interaction, such as electrostatic interaction, but from orbital hybridization arising from overlap in wave functions between MoS$_2$ and hBN. It should be noted that accurate estimation of interlayer distance based on DFT is not straight forward and the degree of the upward shift should be influenced by the accuracy.

Figure 5(b) shows a 2D contour plot of the squared wave function of the VBM of hBN/MoS$_2$/hBN at the Γ-point; the squared wave function was summed over the a-axis. As shown in the figure, there is contribution from the π-band of hBN in the wave function of hBN/MoS$_2$/hBN. A close inspection has revealed that there is a node between MoS$_2$ and hBN, and this means that anti-bonding coupling between wavefunctions of MoS$_2$ and hBN exists in hBN/MoS$_2$/hBN. This anti-bonding coupling leads to the upward shift in VBM at the Γ-valley, resulting in the transformation from a direct gap to an indirect gap. The difference in VB energy of hBN and MoS$_2$ at the Γ-point is ~400 meV, which is much smaller than that in CB, and this relatively small energy difference allows the orbital hybridization between hBN and MoS$_2$. In addition, there is only a vdW gap between the hBN and MoS$_2$ layers, and the wave function of MoS$_2$ at the Γ-point has a significant contribution from Sulphur atoms. Both of them contribute to significant spatial overlap between the wave functions of hBN and MoS$_2$ at the Γ-point, leading to the direct-to-indirect transition.

## Discussion

In contrast to the wavefunction of the VBM at the Γ-point, the wavefunction at the K valley of MoS$_2$ localizes around Molybdenum atoms. This localized nature of the wavefunction plays a major role in small coupling at the K-point; the spatial overlapping between wavefunctions at K-points should be much smaller than that at the Γ-point. In addition, the difference in VB energy between hBN and MoS$_2$ at the K-point strongly depends on relative angle. The smallest energy difference appears when the relative angle is zero, which is not the case in real samples. Consequently, in real samples, the upward shift of VB energy of MoS$_2$ at the K-point should be small, which probably contributes to the observed large $\Delta E$ of 83 meV.

Another scenario that can contribute to the bandstructure change is structural distortion. Due to the vdW force acting between the layers, there is a possibility that encapsulated MoS$_2$ can slightly distort along the direction perpendicular to the 2D plane. The vertical distortion leads to biaxial tensile strain along the direction parallel to the 2D plane, resulting in modification of the band structure (Supplementary Fig. 5). Calculated vertical distortion dependence on the total energy of MoS$_2$ gives Young's modulus of 225 GPa, which means that 225 MPa is needed to induce 0.1 % vertical distortion. Therefore the contribution from distortion cannot be a dominant factor.

## Conclusion

In conclusion, we have revealed that the band structure of MoS$_2$ in hBN/MoS$_2$/hBN is significantly altered by interlayer interaction between hBN and MoS$_2$. Through detailed PL measurements, including the temperature dependence of PL intensity and time-resolved PL intensity, we have shown the existence of dark excitons, whose energy is 83 meV lower than that of the bright excitons. DFT band structure calculation has revealed that the dark excitons should be momentum-forbidden dark excitons, where electrons and holes locate at the K-point and Γ-point, respectively. The existence of the lower-energy momentum-forbidden excitons in pristine monolayer MoS$_2$, originating from the large exciton binding energy, was already pointed out in a previous report. We found, however, that monolayer MoS$_2$ in hBN/MoS$_2$/hBN is not a direct-gap semiconductor anymore due to the orbital hybridization and vdW pressure from encapsulating hBN. This should have an impact on valley-related properties of MoS$_2$, in particular, properties arising from valley-polarized holes.

## Methods
### CVD growth of MoS$_2$ on exfoliated hBN
We have grown monolayer MoS$_2$ crystals onto hBN flakes by CVD method. We used elemental sulfur (Sigma-Aldrich, 99.98 %) and molybdenum oxide (MoO$_3$, Sigma-Aldrich, 99.5%) as precursors for the CVD growth. MoO$_3$/sulfur was placed in/on a quartz tube with an 8.5 mm inner diameter, and the 8.5 mm inner diameter quartz tube was placed in a 26 mm inner diameter quartz tube to avoid unwanted reaction before reaching the hBN flakes. The hBN flakes were prepared on a quartz substrate by the mechanical exfoliation method, and the quartz substrate with hBN flakes was placed in the downstream of the 8.5 mm diameter quartz tube. The quartz tubes were heated with three-zone furnace at 200, 750 and 1100 degree for 20 minutes under Ar flow of 200 sccm; sulfur, MoO$_3$, and a quartz substrate were placed at the coolest, medium and hottest zone, respectively.

### Fabrication of hBN-encapsulated heterostructures
hBN flakes were prepared on a SiO$_2$/Si substrate by the mechanical exfoliation method. One of the hBN flakes on a SiO$_2$/Si was picked up by a PMMA (Microchem A11)/PDMS (Shin-Etsu Silicone KE-106) film on a glass slide, and the picked-up hBN flake was transferred onto a monolayer MoS$_2$ grown on a hBN flake to form a hBN/MoS$_2$/hBN. To achieve sufficient interlayer contact, we have used nano-"squeegee" technique, where AFT tip squeezes out contamination and bubbles from interface between hBN and MoS$_2$.

### Optical measurements
Room temperature PL spectra were measured by using a confocal Raman microscope (Renishaw InVia Raman and Horiba Jobin Yvon LabRAM HR-800) with 488 nm CW laser excitation (COHERENT Sapphire 488 LP). In measurements of temperature dependence of PL spectra and time-resolved PL measurements, we used a home-build microspectroscopy system equipped with a spectrometer (Princeton Instruments IsoPlane SCT320) and a supercontinuum laser system (NKT Photonics SuperK EXTREME); laser beam from the supercontinuum laser was monochromated by a spectrometer (Princeton Instruments SP2150i). In low-temperature measurements, we placed a sample in a cryostat (CryoVac KONTI-Cryostat-Micro) with continuous flowing of liquid $N_2$ under vacuum of ~$10^{-4}$ Pa; CryoVac TIC 304-MA was used to control temperature. Objective lenses (50 ~ 100× and 0.7 ~ 0.85 NA) was used for all measurements.

### First-principles calculations

First-principles density functional theory (DFT) calculations were performed using the Quantum Espresso[28]. Ion-electron interactions were represented by all-electron projector augmented wave potentials[29], and the generalized gradient approximation (GGA) parameterized by Perdew-Burke-Ernzerhof (PBE)[30] were used to account for the electronic exchange and correlation. The wave functions were expanded in a plane wave basis with energy cut-off of 45 Ry. Prior to bandstructure calculation of $MoS_2$ and $hBN/MoS_2/hBN$, the structure of $MoS_2$ was fully relaxed until the components of Hellmann-Feynman forces on the atoms were less than $10^{-5}$ Ry/Å. For bandstructure calculation, the optimized structure of $MoS_2$ was expanded to a 4 x 4 supercell to minimize the lattice mismatch between hBN and $MoS_2$.

### Acknowledgements

This work was supported by JSPS KAKENHI Grant numbers JP16H06331, JP16H03825, JP16H00963, JP15K13283, JP25107002, and JST CREST Grant Number JPMJCR16F3. We thank K. Itami and Y. Miyauchi for sharing the Raman and AFM apparatus. We are grateful to S. Okada for fruitful discussion on ab-initio calculation.

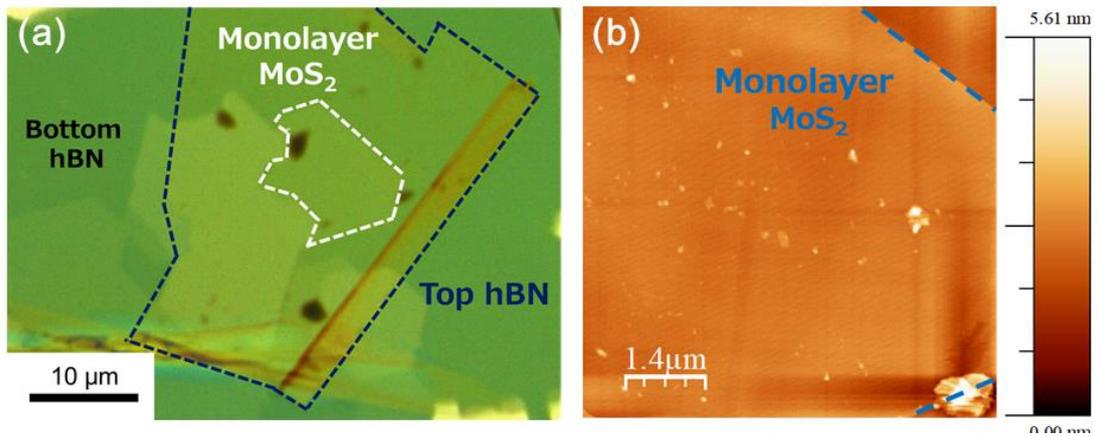

**Figure 1 | Microscopy images of hBN/MoS$_2$/hBN,** (a) an optical image of the hBN/MoS$_2$/hBN prepared. (b) an AFM image of a MoS$_2$ part in the hBN/MoS$_2$/hBN.

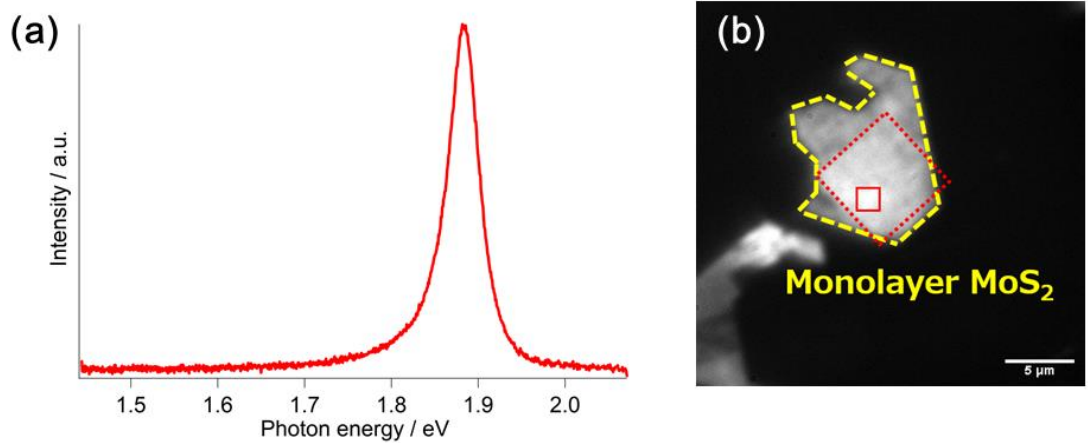

**Figure 2 | A photoluminescence spectrum and image of hBN/MoS$_2$/hBN,** (a) a photoluminescence spectrum of hBN/MoS$_2$/hBN measured at room temperature. Excitation wavelength of 488 nm was used. (b) A photoluminescence image of hBN/MoS$_2$/hBN. Bright part corresponds to encapsulated monolayer MoS$_2$. The red dotted square shows the place, where cleaning with contact-mode AFM was performed. The red square drawn by the solid red line corresponds to the place where integration of PL intensity was performed for evaluation of the temperature dependence of PL intensity.

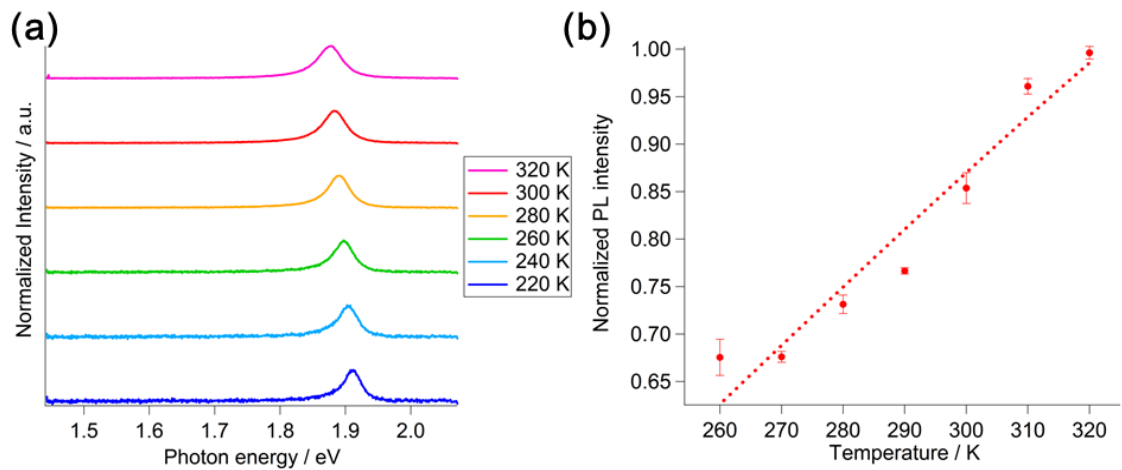

**Figure 3 | Temperature dependence of PL spectra and intensity,** (a) temperature dependence of PL spectra of hBN/MoS$_2$/hBN. All measurements were performed with excitation wavelength of 532 nm. All PL spectra are normalized to their maximum PL intensity. (b) Temperature dependence of PL intensity obtained from PL images. The dotted line corresponds to a fitted line with equation (1).

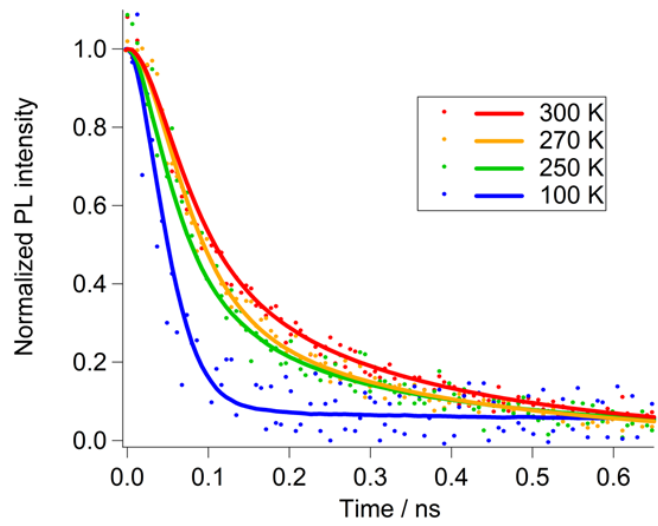

**Figure 4 | Temperature dependence of a time-resolved PL intensity,** A time-resolved PL intensity of hBN/MoS$_2$/hBN measured with time-correlated single photon counting (TCSPC) technique at various temperatures of 100, 250, 270 and 300 K. Dots and solid lines correspond to measured points and fitted curves assuming single exponential decay.

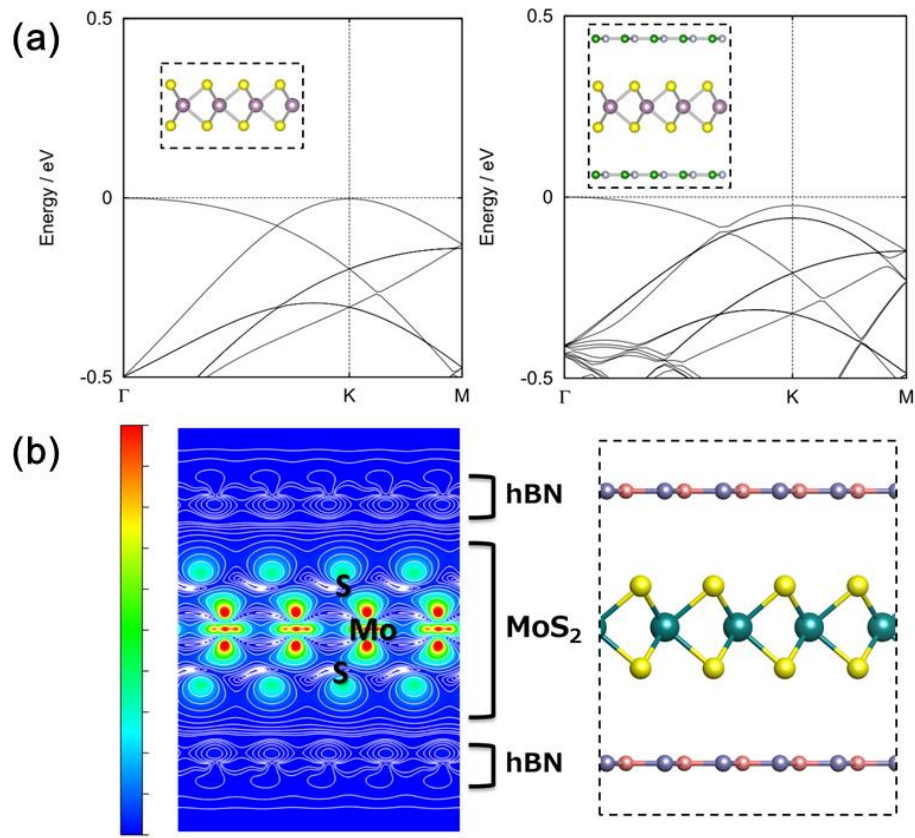

**Figure 5 | Bandstructures of MoS$_2$ and hBN/MoS$_2$/hBN,** Band structures around valence bands of monolayer MoS$_2$ (left) and hBN/MoS$_2$/hBN (left). Insets show structure models used in these calculations, where 4 x 4 and 5 x 5 supercells were used for MoS$_2$ and hBN, respectively, to minimize the lattice mismatch between MoS$_2$ and hBN. (b) Cross section of charge density ($\Psi^2$) of hBN/MoS$_2$/hBN at the Γ-valley projected along the a-axis. The contour lines are drawn in such a way that the differences of charge density at adjacent lines are double. All calculations were performed with Quantum Espresso with energy cut-off of 45 Ry and Monkhorst-Pack k-point mesh of 7 x 7 x 1 were used.

# Supplementary information

## Indirect bandgap of hBN-encapsulated monolayer MoS$_2$

Yosuke Uchiyama[1], Kenji Watanabe[2], Takashi Taniguchi[2], Kana Kojima[3], Takahiko Endo[3], Yasumitsu Miyata[3], Hisanori Shinohara[1] and Ryo Kitaura[1]*

[1]*Department of Chemistry, Nagoya University, Nagoya 464-8602, Japan*

[2]*National Institute for Materials Science, 1-1 Namiki, Tsukuba 305-0044, Japan*
[3]*Department of Physics, Tokyo Metropolitan University, Hachioji, Tokyo 192-0397, Japan*

*Corresponding authors: Ryo Kitaura, r.kitaura@nagoya-u.jp

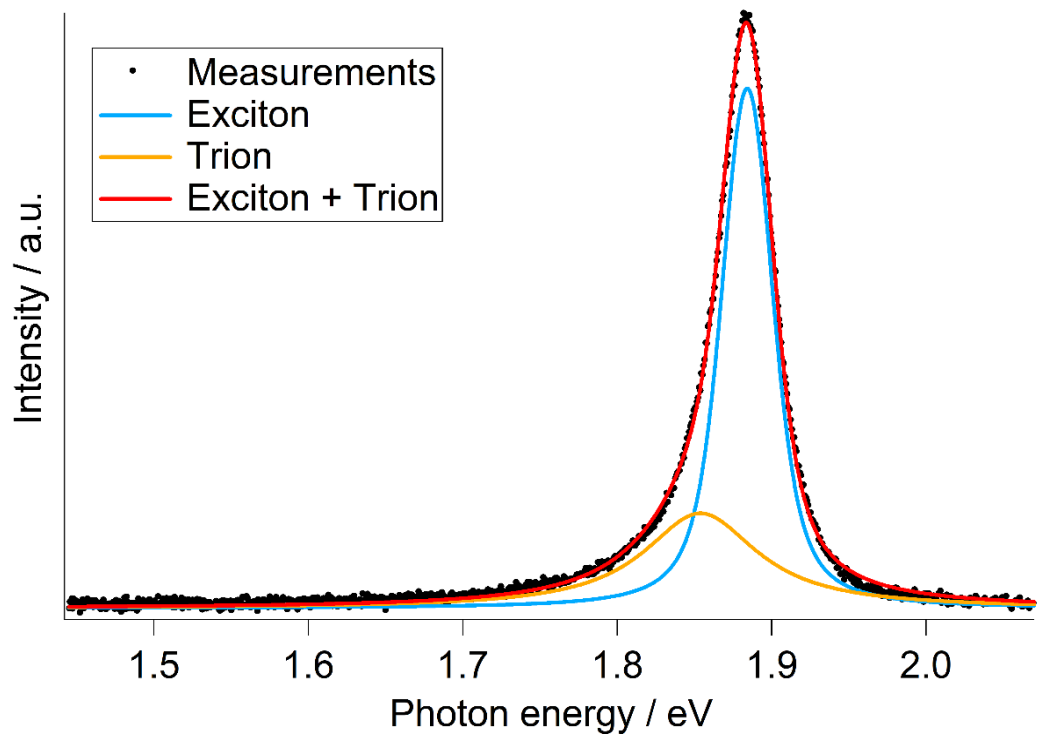

Supplementary Figure 1. A PL spectrum of hBN/MoS$_2$/hBN measured at room temperature.

The black dots are experimental data, which can be decomposed into two contributions, radiative recombination of excitons and trions. As shown in this figure, a least square fitting with two Voigt functions have reproduced the experimental data well, yielding FWHM of 39 and 90 meV for excitons and trions, respectively.

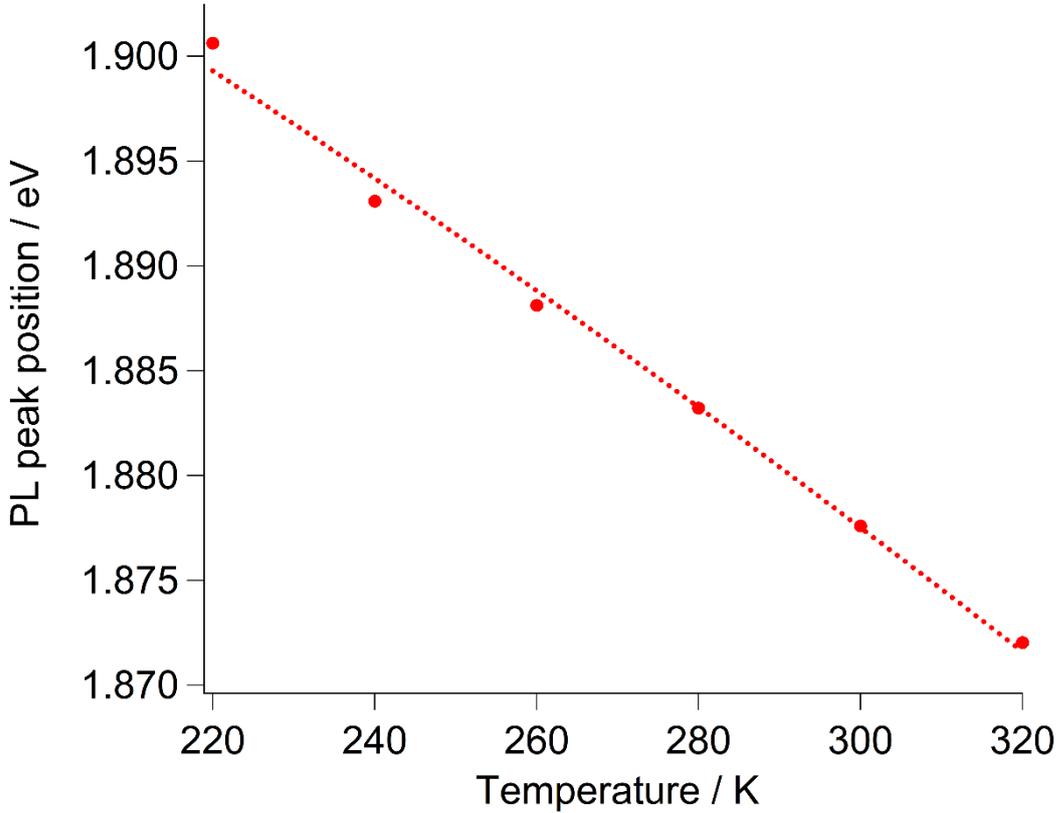

Supplementary Figure 2. Temperature dependence of PL peak positions of exciton emission from hBN/MoS$_2$/hBN.

Temperature dependence of a PL peak position of a semiconductor is usually caused by temperature dependence of bandgap, $E_g(T)$, which can be described by the Varshni's relation shown below.

$$E_g(T) = E_g(0) + \frac{\alpha T^2}{\beta + T}$$

$\alpha$ and $\beta$ in the Varshini's relation are material dependent parameters. A least square fitting using the Varshini's relation reproduces the observed temperature dependence well, giving $\alpha$, $\beta$, and $E_g(0)$ of 0.45 meV/K, 430 K and 1.93 eV, respectively. Because $\alpha$ and $\beta$ is strongly correlated, we fixed $\beta$ to 430 K in the fitting process. Obtained $\alpha$ and $E_g(0)$ are similar to those in the previous report[1].

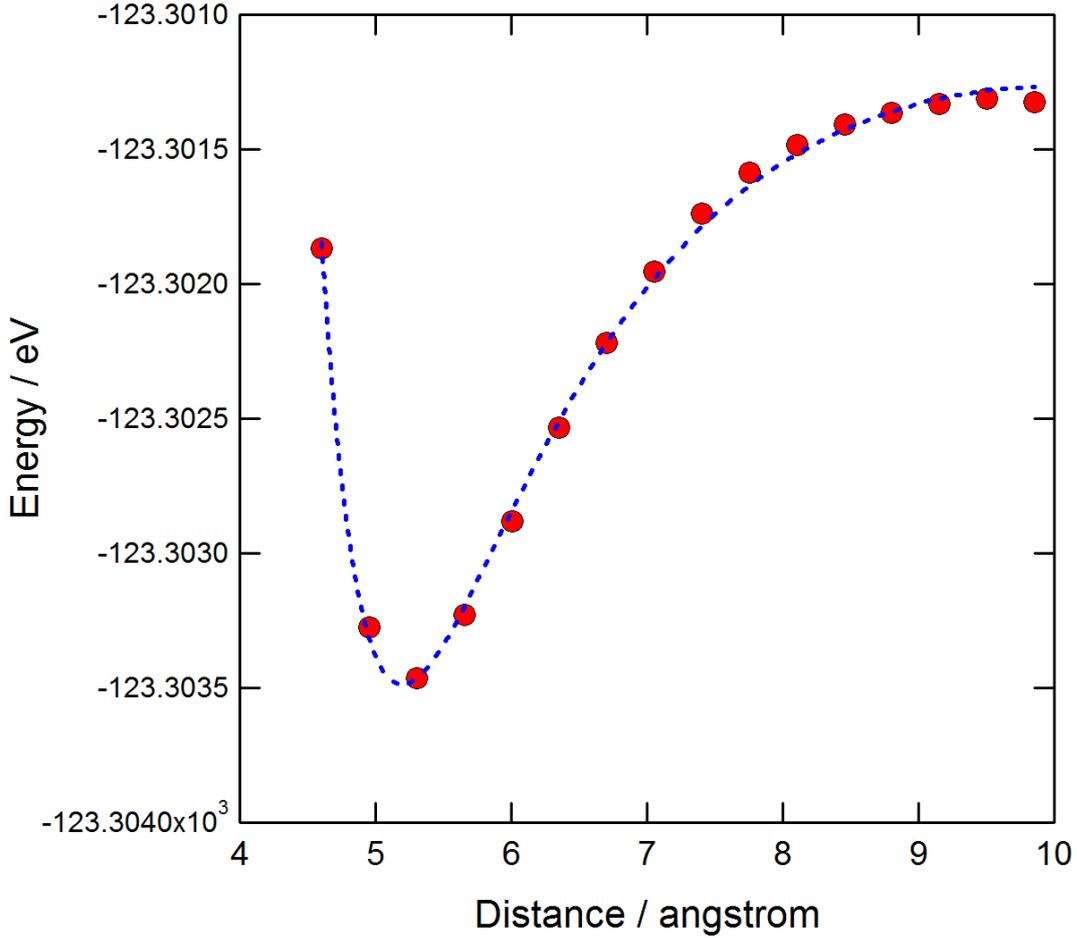

**Supplementary Figure 3. Interlayer distance dependence of total energy of hBN/MoS$_2$/hBN.**

We evaluated interlayer interaction by vdw-DFT implemented in Quantum Espresso[2, 3] with plane-wave basis up to 45 Ry and a 7 x 7 x 1 Monkhorst-Pack k-point mesh. The projector augmented wave method[4] with the Perdew, Burke and Ernzerhof exchange-correlation function[5]. In this calculation, 4 x 4 and 5 x 5 supercells were used for MoS$_2$ and hBN, respectively. Prior to the calculation, full geometry optimization of MoS$_2$ was performed and the optimized structure was used to calculate interlayer interaction. The obtained relation between energy and interlayer distance (difference between z coordinates of Mo and B) was fitted with empirical equation, $E(z) = \varepsilon((\sigma/z)^{d1} - (\sigma/z)^{d2}) + C$. As shown in the figure, a least square fitting with the empirical equation reproduces calculated interlayer dependence of total energies (red points) well, giving parameters of $\varepsilon = 10.9$ eV, $\sigma = 4.50$ Å, d1 = 7.38, d2 = 6.38, C = $-1.2337 \times 10^5$ eV. The obtained optimum interlayer distance, difference in z component of coordinates between Mo and B, is 5.21 Å.

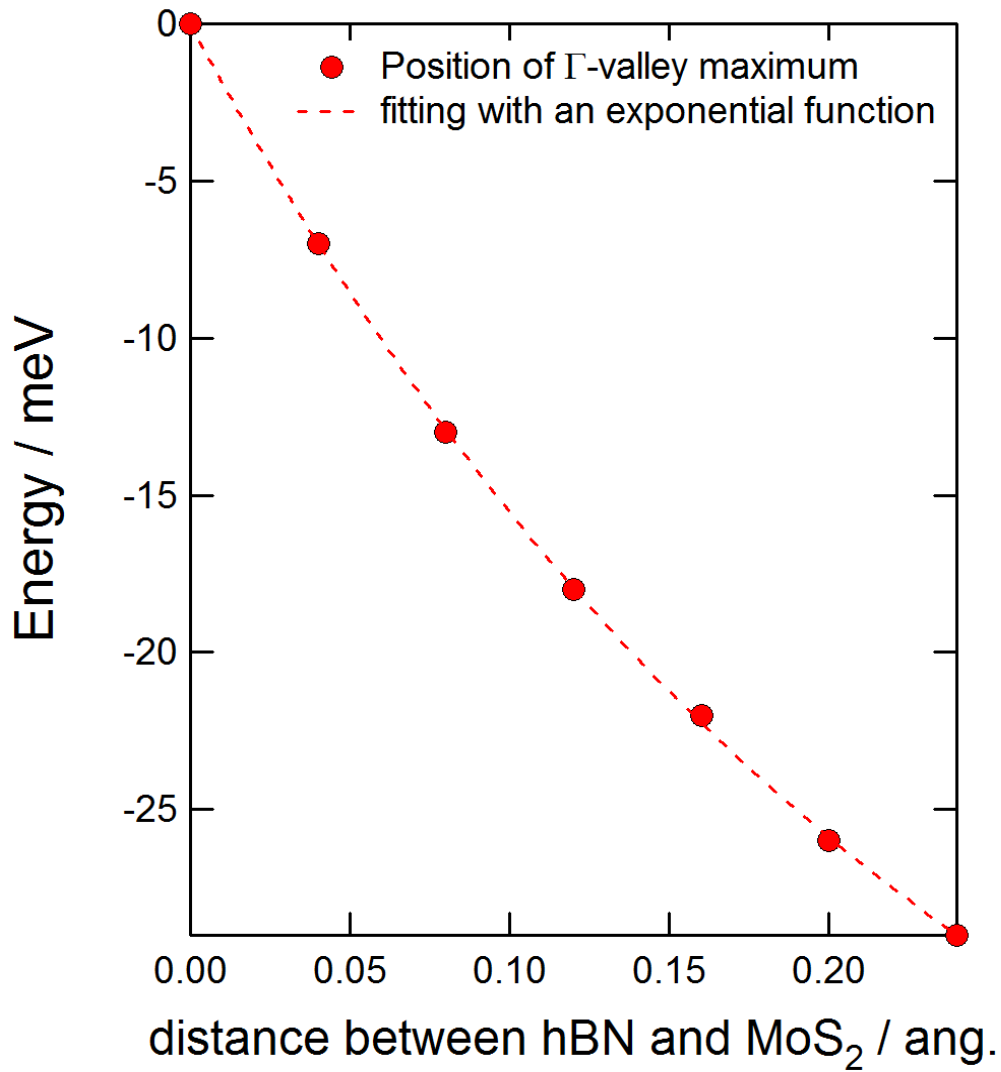

**Supplementary Figure 4. Interlayer distance dependence of position of valence band maximum at the Γ point.**

Positions of Γ-valley maximum are plotted against the interlayer distance. Energy and interlayer distances are respectively shown as relative energy and relative interlayer distance to the optimum interlayer distance determined with vdw-DFT. The interlayer dependence is fitted with $\alpha \exp(-\beta x) + y$, where $\alpha$, $\beta$ and y are fitting parameters and $x$ corresponds to interlayer distance. As seen in the figure, the exponential function reproduces the interlayer dependence well, yielding fitting parameters of $\alpha = 46.647$, $\beta = 4.0406$ and y=-46.676, respectively.

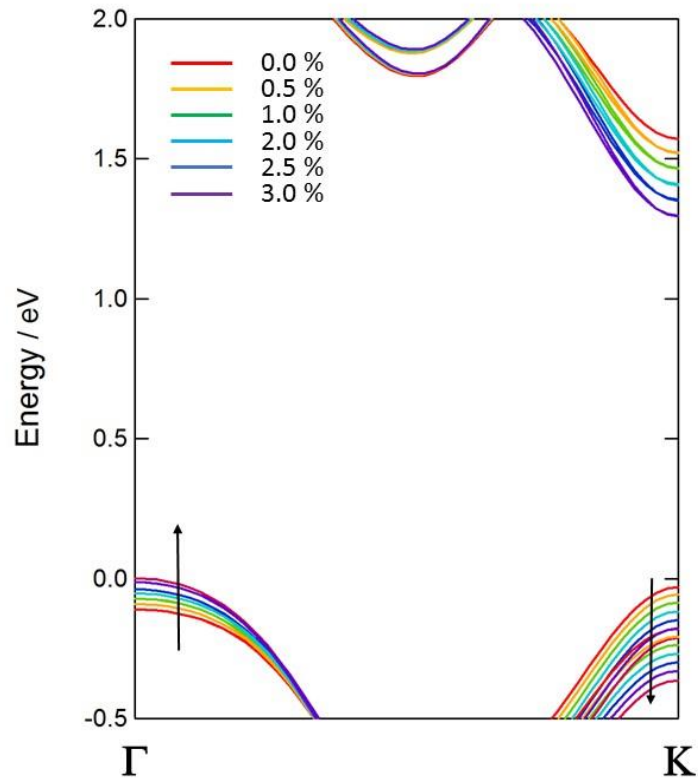

**Supplementary Figure 5. Band structures of monolayer MoS2 with different contractions along the z-axis.**

Band structures of monolayer $MoS_2$ were calculated with different contractions along the z-axis from 0 to 3 %. In this calculation, the length of a-axis was optimized while the z coordinates of Mo and S are fixed to maintain the contraction along the z-axis constant. As seen in the figure, Γ-valley and K-valley show upward and downward shift, respectively, as structure is contracted along the z-axis. All calculations have been performed with VASP[6] with energy cut-off of 400 eV and Monkhorst-Pack k-point mesh of 12 x 12 x 1.

# Supplementary Note 1: Derivatgion of the equation for the temperature dependence of PL intensity

The PL intensity is proportional to the quantum yield (QY), and PL intensity $I(T)$ is given by:

$$I(T) \propto \text{QY} = \frac{\alpha \gamma_{\text{rad}} m_{\text{KK}}}{\alpha \gamma_{\text{rad}} m_{\text{KK}} + \gamma_{\text{dark}} \sum_\mu m_\mu e^{\beta \Delta E_\mu}} \tag{1}$$

where $\gamma_{\text{rad}}$ and $\gamma_{\text{dark}}$ are the radiative and non-radiative decay rate, and $m_{\text{KK}}$, $m_{\text{KK}'}$, $m_{\text{K}\Gamma}$ are the effective mass of the K-K direct, K-K′ spin-forbidden, K-Γ momentum-forbidden excitons, and $\Delta E_\mu$ is the energy difference between the bright state and the dark state, respectively. The denominator and numerator in this equation correspond to total decay rate (sum of non-radiative and radiative decay rate) and radiative decay rate, respectively. The α, which represents the limitation by light cone, is given by:

$$\alpha = 1 - \exp\left(-\frac{\beta \hbar^2 k_{\text{pt}}^2}{2 m_{\text{KK}}}\right) \tag{2}$$

where $\beta$ is the reciprocal number of the product of the Boltzmann constant and the temperature, $\hbar$ is the Dirac constant, and $k_{\text{pt}}$ is the photon momentum, respectively. This equation represents that only excitons in the light cone can couple to photons. The Taylor expansion of this equation is given by:

$$\alpha \sim \beta \frac{E_g}{2 m_{\text{KK}} c^2} \tag{3}$$

where $E_g$ is bandgap, $c$ is the speed of light. The bandgap is calculated from Varshni's relation,

$$E_g(T) = E_g(0) + \frac{\alpha T^2}{\beta + T} \tag{4}$$

where $\alpha$, $\beta$, $E_g(0)$ are the material parameters measured in Supplementary Figure 2. Assume that $\gamma_{\text{dark}} \gg \gamma_{\text{rad}}$, the equation (1) can be simplified to:

$$I(T) \propto \text{QY} \sim \frac{\gamma_{\text{rad}}}{\gamma_{\text{dark}}} \frac{\alpha m_{\text{KK}}}{\sum_\mu m_\mu e^{\beta \Delta E_\mu}} \sim \frac{\alpha m_{\text{KK}}}{\sum_\mu m_\mu e^{\beta \Delta E_\mu}} \tag{5}$$

In this analysis, we consider three states, KK, KK′, KΓ excitons, and equation (5) becomes:

$$I(T) \propto \frac{\alpha m_{\text{KK}}}{m_{\text{KK}} + m_{\text{KK}'} + m_{\text{K}\Gamma} e^{\beta \Delta E_{\text{K}\Gamma}}} \tag{6}$$